\documentclass[conference]{IEEEtran}
\IEEEoverridecommandlockouts
\usepackage{cite}
\usepackage{amsmath,amssymb,amsfonts}
\usepackage{algorithmic}
\usepackage{graphicx}
\usepackage{textcomp}
\usepackage{xcolor}
\def\BibTeX{{\rm B\kern-.05em{\sc i\kern-.025em b}\kern-.08em
    T\kern-.1667em\lower.7ex\hbox{E}\kern-.125emX}}

\usepackage{algorithm}
\usepackage{tikz}	
\usetikzlibrary{backgrounds,fit,arrows,decorations.pathreplacing,positioning}
\usepackage[normalem]{ulem}	

\DeclareMathOperator{\sgn}{sign}
\DeclareMathOperator*{\argmax}{arg\,max}

\newcommand{\dlt}{\delta}

\newcommand{\La}{\Lambda}

\newcommand{\teq}{\triangleq}

\newcommand{\sM}{{\cal M}}
\newcommand{\sN}{{\cal N}}

\newcommand{\mbf}[1]{\pmb{#1}}

\newcommand{\ps}[1]{\left( {#1} \right)}

\newcommand{\beq}[1]{$${ #1 }$$}
\newcommand{\beql}[2]{\begin{equation}\label{#1}{ #2 }\end{equation}}
\newcommand{\beqs}[1]{\begin{align*} #1 \end{align*}}
\newcommand{\beqsl}[2]{\begin{equation}\label{#1}\begin{aligned} #2 \end{aligned}\end{equation}}
\newcommand{\bmtx}[1]{\left[ \begin{matrix}  #1  \end{matrix} \right]}
\newcommand{\bsmtx}[1]{\left[ \begin{smallmatrix} #1 \end{smallmatrix} \right]}

\begin{document}

\title{Refined Belief-Propagation Decoding of Quantum Codes with Scalar Messages
}

\author{\IEEEauthorblockN{Kao-Yueh~Kuo and Ching-Yi~Lai}
\IEEEauthorblockA{\textit{Institute of Communications Engineering, National Chiao Tung University} \\
Hsinchu 30010, Taiwan \\
kykuo@nctu.edu.tw and cylai@nctu.edu.tw}
}

\maketitle

\begin{abstract}
Codes based on sparse matrices have good performance and can be efficiently decoded by belief-propagation (BP).
Decoding binary stabilizer codes needs a quaternary BP for (additive) codes over GF(4), 
	which has a higher check-node complexity compared to a binary BP for codes over GF(2).
	Moreover, BP decoding of stabilizer codes suffers a performance loss from the short cycles in the underlying Tanner graph.
In this paper, we propose a refined BP algorithm for decoding quantum codes by passing scalar messages. 
	For a given error syndrome, this algorithm decodes to the same output as the conventional quaternary BP but with a check-node complexity the same as binary BP.
	As every message is a scalar, the message normalization can be naturally applied to improve the performance.
	Another observation is that the message-update schedule affects the BP decoding performance against short cycles.
We show that running BP with message normalization according to a serial schedule (or other schedules) may significantly improve the
decoding performance and error-floor in computer simulation.
\end{abstract}

\begin{IEEEkeywords}
Quantum stabilizer codes, LDPC codes, sparse matrices, belief-propagation, sum-product algorithm, message-update schedule, message normalization.
\end{IEEEkeywords}

\section{Introduction} \label{sec:Intro}

In the 1960s, Gallager proposed the low-density parity-check (LDPC) codes (or called sparse codes) 
	and sum-product (decoding) algorithm to have a near Shannon-capacity performance for many channels, including
	the binary symmetric channel (BSC) and additive white Gaussian noise (AWGN) channel \cite{Gal62,Gal63,MN95,MN96,Mac99}. 
	The sum-product algorithm runs an iterative message-passing 
	on the Tanner graph \cite{Tan81} defined by a parity-check matrix of the code \cite{Wib96,AM00,KFL01},
where the procedure is usually understood as a realization of Pearl's belief-propagation (BP)---a widely-used algorithm in machine-learning and artificial-intelligence \cite{Pea88,MMC98,YFW03}.

The parity-check matrix of a classical code is used to do syndrome measurements for (syndrome-based) decoding. 
In the 1990s, quantum error-correction was proposed to be done in a similar way by using stabilizer codes \cite{Shor95,GotPhD,CS96,Steane96,CRSS98}. 
There will be a check matrix defined for a stabilizer code to do the projective (syndrome) measurements for decoding.
Many sparse stabilizer codes have been proposed, such as topological codes \cite{Kit03,BM06}, random sparse codes (in particular, bicycle codes) \cite{MMM04}, and hypergraph-product codes \cite{TZ14}. 

Decoding a binary stabilizer code is equivalent to decoding an {additive} code over GF($q=4$) (representing Pauli errors $I,X,Y,Z$) \cite{CRSS98}. 
That is, a quaternary BP (BP$_4$) is required, with every message conventionally a length-four vector to represent a distribution over GF(4). 
In contrast, decoding a binary classical code only needs a binary BP (BP$_2$), with every message a scalar (e.g., as a \emph{log-likelihood ratio} (LLR) \cite{Gal63} or \emph{likelihood difference} (LD) \cite{Mac99}).
As a result, BP$_2$ has a check-node efficiency $q^2=16$ times better than the conventional BP$_4$ \cite{DM98}.\footnote{ 
	The complexity can be $O(q\log q)$ for the fast Fourier transform (FFT) method \cite{MD01,DF07}. 
	But a comparison based on this is more suitable for a larger $q>4$, since there is additional cost running FFT and inverse FFT.
	}
To reduce the complexity, it often treats a binary stabilizer code as a binary classical code with doubled length so that BP$_2$ can be used \cite{MMM04}, followed by additional processes to handle the $X$/$Z$ correlations \cite{DT14,ROJ19}.

A stabilizer code inevitably has many four-cycles in its Tanner graph, which degrade the performance of BP \cite{MMM04,PC08}.
To improve, additional pre/post-processes are usually used, 
	such as heuristic flipping from nonzero syndrome bits \cite{PC08}, 
	(modified) enhanced-feedback \cite{Wan+12,Bab+15}, training a neural BP \cite{LP19}, 
	augmented decoder (adding redundant rows to the check matrix) \cite{ROJ19}, and ordered statistics decoding (OSD) \cite{PK19}.

In this paper, we first simplify the conventional BP$_4$.
An important observation is that, though a binary stabilizer code is considered as a quaternary code, its error syndrome is binary.
Inspired by this observation and MacKay's LD-based computation rule (\mbox{$\delta$-rule}) (cf. (47)--(53) in \cite{Mac99}), we derive a \mbox{$\delta$-rule} based BP$_4$ that is equivalent to the conventional~BP$_4$ but~passing every message as a scalar, which improves the check-node efficiency 16 times and does not need additional processes to handle the $X$/$Z$ correlations.


Then we improve BP$_4$ by introducing classical BP techniques. 
First, the messages can be updated in different orders (schedules) \cite{ZF05,SLG07,GH08}.
Second, as we have scalar messages, the message normalization can be applied \cite{CF02b,CDE+05,YHB04}.
%
With these techniques, our (scalar-based) BP$_4$ shows a much improved empirical performance in computer simulation.

This paper is organized as follows. 
In Sec.\,\ref{sec:ClsBP}, we define the notation and review BP$_2$. 
In Sec.\,\ref{sec:QuanBP}, we show that BP$_4$ 
	can be efficiently done by passing scalar messages.  
In Sec.\,\ref{sec:NorQBP}, we introduce the message normalization and provide simulation results.
Finally, we conclude in Sec.\,\ref{sec:Conclu}. 

\section{Classical Binary Belief-Propagation Decoding} \label{sec:ClsBP}

\subsection{The notation and parallel/serial BP$_2$}

We consider the syndrome-based decoding. 
Consider an $[N,K]$ classical binary linear code defined by an \mbox{$M \times N$} parity-check matrix $H\in\{0,1\}^{M\times N}$ (not necessarily of full rank) with $M\geq N-K$.
Upon the reception of a noisy codeword disturbed by an unknown error $E=(E_1,E_2,\dots, E_N) \in \{0,1\}^N$, 
	syndrome measurements (defined by rows of $H$) are performed to generate a syndrome vector $z\in\{0,1\}^M$.
%
%
The decoder is given $H$, $z$, and an a priori distribution $\mbf p_n = (p_n^{(0)},p_n^{(1)})$ for each $E_n$ to infer an $\hat E$ such that the probability \mbox{$P(\hat E = E)$} is as high as possible.\footnote{
	Giving $\mbf p_n$ can be according to a channel model \cite{Mac99} or even a fixed distribution (useful if the channel parameter is unknown) \cite{HFI12}. Here, we uses the former case since it usually has a better performance.
	}
BP decoding is done by updating each $\mbf p_n$ as an a posterior distribution $\mbf q_n = (q_n^{(0)},q_n^{(1)})$ 
	and inferring $\hat E = (\hat E_1,\hat E_2,\dots,\hat E_N)$ by $\hat E_n = \argmax_{b\in\{0,1\}} q_n^{(b)}$.
This is done by doing an iterative {\it message-passing} on the {\it Tanner graph} defined by $H$.

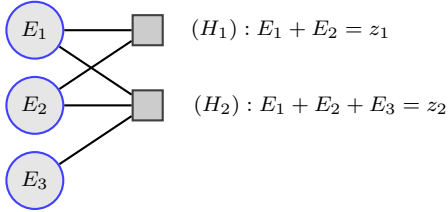
\begin{figure}[h!] \centering ~~~~~~~~ \begin{tikzpicture}[node distance=1.3cm,>=stealth',bend angle=45,auto]

\tikzstyle{chk}=[rectangle,thick,draw=black!75,fill=black!20,minimum size=4mm]
\tikzstyle{var}=[circle,thick,draw=blue!75,fill=gray!20,minimum size=4mm,font=\footnotesize]
\tikzstyle{VAR}=[circle,thick,draw=blue!75,fill=blue!20,minimum size=5mm,font=\footnotesize]
\tikzstyle{fac}=[anchor=west,font=\footnotesize]

\node[var] (x3) at (0,0) {$E_3$};
\node[var] (x2) at (0,1) {$E_2$};
\node[var] (x1) at (0,2) {$E_1$};
\node[chk] (c1) at (1.5,2) {};
\node[chk] (c2) at (1.5,1) {};

\draw[thick] (x1) -- (c1) -- (x2);
\draw[thick] (x1) -- (c2) -- (x3);
\draw[thick] (x2) -- (c2);


\node[fac] [right of=c1,xshift=6mm] {$(H_1): E_1+E_2=z_1$};
\node[fac] [right of=c2,xshift=10mm] {$(H_2): E_1+E_2+E_3=z_2$};


\end{tikzpicture}
	\caption[Roots associated to the Cartan matrix]{
		The Tanner graph of an example $H=\left[\begin{smallmatrix}1 &1 &0\\ 1 &1 &1\end{smallmatrix}\right]$. 
		The three circles are variable nodes and the two squares are check nodes.
	} \label{fig:H2x3}
\end{figure}

The Tanner graph corresponding to $H$ is a bipartite graph consisting of $N$ variable nodes and $M$ check nodes, 
and there is an edge $(m,n)$ connecting check node $m$ and variable node~$n$ if entry $H_{mn}=1$.
Let $H_m$ be row $m$ of $H$.
An example  Tanner graph of $H=\left[\begin{smallmatrix}1 &1 &0\\ 1 &1 &1\end{smallmatrix}\right]$ is shown in Fig.\,\ref{fig:H2x3}. 

For the iterative message-passing, there will be a variable-to-check message $d_{n\to m}$ and a check-to-variable message $\delta_{m\to n}$ passed on edge $(m,n)$.
%
%
Let $\sN(m) = \{n: H_{mn}=1\}$ and $\sM(n) = \{m: H_{mn}=1\}$.
(And we will write things like $\sM(n)\setminus\{m\}$ as $\sM(n)\setminus m$ to simplify the notation.)
Assume that every $\delta_{m\to n}=0$ initially. An iteration is done by:
\begin{itemize}\setlength{\itemindent}{-1em}
\item $d_{n\to m}$ is computed by \mbox{$\{\delta_{m'\to n}: m'\in\sM(n)\setminus m\}$} with~$\mbf p_n$,  
\item $\delta_{m\to n}$ is computed by \mbox{$\{d_{n'\to m}: n'\in\sN(m)\setminus n\}$} with $z_m$,  
\item $\mbf{q}_n$ is computed by \mbox{$\{\delta_{m\to n}: m\in\sM(n)\}$} with~$\mbf p_n$.
\end{itemize}
%
The messages $d_{n\to m}$ and $\delta_{m\to n}$  will be denoted, respectively, by $d_{mn}$ and $\delta_{mn}$ for simplicity.
%
The message-passing is usually done with a parallel schedule, referred to as {\it parallel BP$_2$} (\mbox{cf. \cite[Algorithm~1]{KL20}},
	which has $d_{mn}$ and $\delta_{mn}$ as {\it likelihood differences} (LDs) and is due to MacKay and Neal \cite{MN95,MN96,Mac99}).

The message-passing can also be done with a serial schedule, referred to as {\it serial BP$_2$} \cite[Algorithm~2]{KL20}.
	We compared parallel/serial BP$_2$ by decoding a $[13298,3296]$ code in \cite[Figs.\,4 and 5]{KL20},
	showing that serial BP$_2$ has a better convergence within limited iterations (as expected in \cite{ZF05,SLG07,GH08}).

\subsection{The simplified rules for BP$_2$} \label{sec:Simp_BP2}

The computation in parallel/serial BP$_2$ can be substantially simplified and represented by simple update rules,
	corresponding to \mbox{weight-2} and \mbox{weight-3} rows.
	These simple rules can in~turn be extended to define the general rules \mbox{\cite[Sec.\,V-E]{KFL01}}. 
%
Consider $H=\bsmtx{1&1&0\\ 1&1&1}$ as in Fig.\,\ref{fig:H2x3}. 
Recall that we denote $P(E_1=0)=p_1^{(0)}$ and $P(E_1=1)=p_1^{(1)}$.
Let $E_2+E_3$ be a super bit indexed by $\{2,3\}$.
Then we have the representation $P(E_2+E_3 = 0)=p_{\{2,3\}}^{(0)}$ and $P(E_2+E_3 = 1)=p_{\{2,3\}}^{(1)}$.
By Bayes rule and some derivations, we can update the (conditional) distribution of $E_n$, say $n=1$, by:

\subsubsection*{LD-rule ($\delta$-rule)}
The distribution of $E_1$ can be updated by (conditions) $H_1$ and $H_2$, respectively, by:
\beqsl{eq:dH1}{	&H_1=[1,1,0]:~\bmtx{p_1^{(0)}\\p_1^{(1)}} \to \bmtx{q_1^{(0)}\\q_1^{(1)}} \propto \bmtx{p_1^{(0)}p_2^{(0)}\\ p_1^{(1)}p_2^{(1)}},\\
				&H_2=[1,1,1]:~\bmtx{p_1^{(0)}\\p_1^{(1)}} \to \bmtx{q_1^{(0)}\\q_1^{(1)}} \propto \bmtx{p_1^{(0)}p_{\{2,3\}}^{(0)}\\ p_1^{(1)}p_{\{2,3\}}^{(1)}},
}
where 
\beqsl{eq:dH2}{
	p_{\{2,3\}}^{(0)} &= p_2^{(0)}p_3^{(0)}+p_2^{(1)}p_3^{(1)} = (1+d_2d_3)/2, \\
	p_{\{2,3\}}^{(1)} &= p_2^{(0)}p_3^{(1)}+p_2^{(1)}p_3^{(0)} = (1-d_2d_3)/2, 
}
where $d_n$ is the {\it likelihood difference} (LD) of $E_n$,
\beql{eq:dH3}{  d_n = p_n^{(0)}-p_n^{(1)}.  }
We describe the $\delta$-rule following \cite{Mac99} rather than \cite{KFL01}.
This allows us to generalize the computation to the quantum case later.
The $\delta$-rule is equivalent to the LLR-rule as follows.

\subsubsection*{LLR-rule ($\Lambda$-rule)}
Let $\Lambda_n=\ln\frac{p_n^{(0)}}{p_n^{(1)}}$ be the {\it log-likelihood ratio} (LLR) of $E_n$.
Define $f(x) = \ln\frac{e^x+1}{e^x-1} = \ln\ps{ \coth(\frac{x}{2}) }$ like what Gallager did~\cite{Gal62}, where $x\ge 0$, $f(0)\teq \infty$, and $f(\infty)\teq 0$. 
(Note that $f^{-1} = f$.)
The LLR of $E_1$ can be updated by (conditions) $H_1$ and $H_2$, respectively, by:
\beqs{	&H_1=[1,1,0]:~\La_{1} \to \La_{1} + \La_2, 	\\		
		&H_2=[1,1,1]:~\La_{1} \to \La_{1} + \La_{\{2,3\}} = \La_1 + (\La_{2}\boxplus\La_{3}), \nonumber
}
where $\La_{\{2,3\}} = \La_{2}\boxplus\La_{3}$ is defined by 
\begin{equation}	\La_{2}\boxplus\La_{3} = \sgn(\La_2\La_3)\, f\ps{f(|\La_2|) + f(|\La_3|)}.	\label{bplus1} \end{equation}
It only needs an addition after a transform by~$f$. 
%
To prevent handling the signs separately, it can use a transform by $\tanh$ but needs a multiplication \cite{HOP96} by
	\beqsl{bplus2}{	\La_{2}\boxplus\La_{3} &= 2\tanh^{-1}\ps{ \tanh(\La_2/2) \tanh(\La_3/2) }. }

Using simple algebra can confirm that the $\La_{2}\boxplus\La_{3}$ in \eqref{bplus1} and \eqref{bplus2} are equivalent, 
and $\La_{2}\boxplus\La_{3} = \ln\left( p_{\{2,3\}}^{(0)} / p_{\{2,3\}}^{(1)} \right)$ for the $p_{\{2,3\}}^{(0)}$ and $p_{\{2,3\}}^{(1)}$ defined in \eqref{eq:dH2}. 
More equivalent rules can be found in~\cite[Sec.\,V-E]{KFL01}.

Depending on the application (the channel model, decoder hardware, etc.), a suitable rule can be chosen.
For example, the $\Lambda$-rule is suitable for the AWGN channel (since the received symbol's magnitude is linearly mapped to~$|\La_n|$ \cite{Gal63}), and the operation $\boxplus$ can be efficiently computed/approximated (e.g., by table lookup \cite{CDE+05}).
On the other hand, the $\delta$-rule is suitable for the BSC, and the computation can be efficiently programmed/computed by a general-purpose computer.

The $\delta$-rule is very suitable for our purpose of decoding quantum codes in the next section.

\section{Quaternary BP Decoding for Quantum Codes} \label{sec:QuanBP}

\subsection{The simplified rule for BP$_4$} \label{sec:Simp_BP4}

We will use the notation based on Pauli operators
\mbox{$\left\{	I=\left[\begin{smallmatrix}1 &0\\0&1\end{smallmatrix}\right], 
				X=\left[\begin{smallmatrix}0 &1\\1&0\end{smallmatrix}\right], 
				Y=\left[\begin{smallmatrix}0 &-i\\i&0\end{smallmatrix}\right], 
				Z=\left[\begin{smallmatrix}1 &0\\0&-1\end{smallmatrix}\right]\right\}$} \cite{NC00},
and consider $N$-fold Pauli operators with an inner product 
\begin{equation} \label{eq:in_prod}
\langle \cdot,\cdot\rangle: \{I,X,Y,Z\}^{\otimes N}\times \{I,X,Y,Z\}^{\otimes N} \rightarrow \{0,1\},
\end{equation}
which has an output 0 if the inputs commute and an output~1 if the inputs anticommute.
(This is mathematically equivalent to mapping $\{I,X,Y,Z\}^{\otimes N} \to$ GF(4)$^N$ with a Hermitian trace inner-product to the ground field GF(2) \cite{CRSS98}).
It suffices to omit $\otimes$ in the following discussion. 
An unknown error will be denoted by $E=E_1E_2\cdots E_N\in\{I,X,Y,Z\}^N$.

There will a check matrix $S\in\{I,X,Y,Z\}^{M\times N}$ used to generate a syndrome $z\in\{0,1\}^M$.
Consider a simple example 
$S = \left[\begin{smallmatrix}
S_{11} & S_{12} & I \\
S_{21} & S_{22} & S_{23} \\
\end{smallmatrix}\right]$,
where there are five non-identity entries. 
For simplicity, fix it as 
$S = \left[\begin{smallmatrix}
X & Y & I \\
Z & Z & Y \\
\end{smallmatrix}\right]$
and the other cases can be handled similarly.
The Tanner graph (with generalized edge types $X,Y,Z$) of the example $S$ is shown in Fig.\,\ref{fig:S2x3}.
\begin{figure}[h] \centering ~~~~ \begin{tikzpicture}[node distance=1.3cm,>=stealth',bend angle=45,auto]

\tikzstyle{chk}=[rectangle,thick,draw=black!75,fill=black!20,minimum size=4mm]
\tikzstyle{var}=[circle,thick,draw=blue!75,fill=gray!20,minimum size=4mm,font=\footnotesize]
\tikzstyle{VAR}=[circle,thick,draw=blue!75,fill=blue!20,minimum size=5mm,font=\footnotesize]
\tikzstyle{fac}=[anchor=west,font=\footnotesize]

\node[var] (x3) at (0,0) {$E_3$};
\node[var] (x2) at (0,1) {$E_2$};
\node[var] (x1) at (0,2) {$E_1$};
\node[chk] (c1) at (1.5,2) {};
\node[chk] (c2) at (1.5,1) {};

\draw[thick] (x1) -- (c1);
\draw[thick,dashed] (x2) -- (c1);
\draw[thick,densely dotted] (x1) -- (c2) -- (x2);
\draw[thick,dashed] (x3) -- (c2);


\node[fac] [right of=c1,xshift=12mm] {$(S_1):~\langle E_1,X\rangle+\langle E_2,Y\rangle = z_1$};
\node[fac] [right of=c2,xshift=19mm] {$(S_2):~\langle E_1,Z\rangle+\langle E_2,Z\rangle+\langle E_3,Y\rangle = z_2$};

\node[fac] (Xl) [right of=x3,xshift=5mm,yshift=7] {};
\node[fac] (Xr) [right of=x3,xshift=20mm,yshift=7] {$X$};
\draw[thick] (Xl) -- (Xr);
\node[fac] (Yl) [right of=x3,xshift=5mm,yshift=0] {};
\node[fac] (Yr) [right of=x3,xshift=20mm,yshift=0] {$Y$};
\draw[thick,dashed] (Yl) -- (Yr);
\node[fac] (Zl) [right of=x3,xshift=5mm,yshift=-7] {};
\node[fac] (Zr) [right of=x3,xshift=20mm,yshift=-7] {$Z$};
\draw[thick,densely dotted] (Zl) -- (Zr);

\end{tikzpicture}
	\caption[Roots associated to the Cartan matrix]{
		The Tanner graph of of the example $S = \left[\begin{smallmatrix} X & Y & I \\ Z & Z & Y \\\end{smallmatrix}\right]$.
	} \label{fig:S2x3}
\end{figure}

For decoding quantum codes, we are to update an initial distribution $\mbf p_n=(p_n^I,p_n^X,p_n^Y,p_n^Z)$ of $E_n$ as an updated (conditional) distribution $\mbf q_n=(q_n^I,q_n^X,q_n^Y,q_n^Z)$. 
Again (as in Sec.\,\ref{sec:Simp_BP2}) assume that the syndrome is a zero vector and we are to update the distribution of $E_n$ with $n=1$. 
By Bayes rule and some derivations, we have: 
%
\subsubsection*{LD-rule ($\delta$-rule) for BP$_4$} 
The distribution of $E_1$ can be updated by (conditions) $S_1$ and $S_2$, respectively, by:
	\allowdisplaybreaks
	\beqsl{eq:dS1}{
	&S_1=XYI:~\bmtx{p_1^I\\ p_1^X\\ p_1^Y\\ p_1^Z} \to \bmtx{q_1^I\\ q_1^X\\ q_1^Y\\ q_1^Z} \propto
		\bmtx{p_1^I\,(p_2^I+p_2^Y)\\ p_1^X\,(p_2^I+p_2^Y)\\ p_1^Y\,(p_2^X+p_2^Z)\\ p_1^Z\,(p_2^X+p_2^Z)},\\
	&S_2=ZZY:~\bmtx{p_1^I\\ p_1^X\\ p_1^Y \\p_1^Z} \to \bmtx{q_1^I\\ q_1^X\\ q_1^Y\\ q_1^Z} \propto
		\bmtx{p_1^I\,p_{\{2,3\}}^{(0)}\\ p_1^X\,p_{\{2,3\}}^{(1)}\\ p_1^Y\,p_{\{2,3\}}^{(1)}\\ p_1^Z\,p_{\{2,3\}}^{(0)}},
	}
where (since $S_{22}S_{23} = ZY$)
\beqs{
	p_{\{2,3\}}^{(0)} &= p_2^Ip_3^I+p_2^Ip_3^Y+p_2^Xp_3^X+p_2^Xp_3^Z \\		
			&\qquad +p_2^Zp_3^I+p_2^Zp_3^Y+p_2^Yp_3^X+p_2^Yp_3^Z, \\
	p_{\{2,3\}}^{(1)} &= p_2^Ip_3^X+p_2^Ip_3^Z+p_2^Xp_3^I+p_2^Xp_3^Y \\		
			&\qquad +p_2^Zp_3^X+p_2^Zp_3^Z+p_2^Yp_3^I+p_2^Yp_3^Y.
}
The last two probabilities (with 16 multiplications as above) can be efficiently computed (with one multiplication) by
\beqsl{eq:dS2}{p_{\{2,3\}}^{(0)} = \frac{1+\dlt_2\dlt_3}{2} \quad\text{and}\quad
				p_{\{2,3\}}^{(1)} = \frac{1-\dlt_2\dlt_3}{2},
}
where (since $S_{22}=Z$ and $S_{23}=Y$)
\beqsl{eq:dS3}{	\dlt_2 &= (p_2^I+p_2^Z) - (p_2^X+p_2^Y), \\ 
				\dlt_3 &= (p_3^I+p_3^Y) - (p_3^X+p_3^Z).
}

Observe the similarity and difference in \eqref{eq:dH1}--\eqref{eq:dH3} and \mbox{\eqref{eq:dS1}--\eqref{eq:dS3}}.
This even suggests a way to generalize the above strategy to a case of higher-order GF($q=2^l$) with binary syndromes.

This pre-add strategy is like a partial FFT-method since for GF($q=2^l$), 
	the basic operation of the ground filed GF(2) is addition (subtraction)~\cite{DF07}.  
However, the FFT-method needs $O(q\log q)$ additions and $O(q)$ multiplications to combine two distributions \cite{MD01}.
The above strategy, by generalizing \eqref{eq:dS2} and~\eqref{eq:dS3}, would only need $O(q)$ additions and $O(1)$ multiplication to combine two distributions
	if the syndrome is binary.
Another point is that the FFT-method still treats a message as a vector over GF($q$), 
but we treat the message as a scalar and this opens new possibilities for improvement
(e.g., using message normalization (shown later) or allowing a neural BP$_4$, where (since training needs scalar messages) a previous neural~BP is only considered with BP$_2$ without $X$/$Z$ correlations \cite{LP19}).

\subsection{The conventional BP$_4$ with vector messages}

An $[[N,K]]$ stabilizer code is a $2^K$-dimensional subspace of $\mathbb{C}^{2^N}$
fixed by the operators (not necessarily all independent) defined by a check matrix $S\in\{I,X,Y,Z\}^{M\times N}$ with $M\geq N-K$. 
Each row $S_m$ of $S$ corresponds to an \mbox{$N$-fold} Pauli operator that stabilizes the code space. 
The matrix~$S$ is self-orthogonal with respect to the inner product \eqref{eq:in_prod}, i.e., $\langle S_m,S_{m'}\rangle=0$ for any two rows $S_m$ and $S_{m'}$.
The code space is the joint-($+1$) eigenspace of the rows of $S$.
The vectors in the (multiplicative) rowspace of $S$ are  called \emph{stabilizers} \cite{NC00}.

It suffices to consider discrete errors like Pauli errors per the error discretization theorem \cite{NC00}.
Upon the reception of a noisy coded state suffered from an unknown $N$-qubit error $E\in\{I,X,Y,Z\}^N$,
$M$ stabilizers $\{S_m\}_{m=1}^M$ are measured to determine a \emph{binary error syndrome} $z = (z_1,z_2,\dots,z_M)\in\{0,1\}^M$ using \eqref{eq:in_prod}, 
where 
\begin{align}
z_m= \langle E,S_m \rangle \in\{0,1\}. \label{eq:syndrome_bit}
\end{align}
Given $S$, $z$, and an a priori distribution ${\mbf p}_n= (p_n^I,p_n^X,p_n^Y,p_n^Z)$ for each $E_n$ 
(e.g., let ${\mbf p}_n=(1-\epsilon,\, \epsilon/3,\, \epsilon/3,\, \epsilon/3)$ for a depolarizing channel with error rate $\epsilon$),
a decoder has to estimate an $\hat E\in\{I,X,Y,Z\}^N$ such that $\hat E$ is equivalent to $E$, up to a stabilizer, 
with a probability as high as possible. 
(The solution is not unique due to the degeneracy \cite{PC08,KL13_20}.)

For decoding quantum codes, the neighboring sets are $\sN(m)=\{n:S_{mn}\ne I\}$ and $\sM(n)=\{m:S_{mn}\ne I\}$. 
Let $E|_{\sN(m)}$ be the restriction of  $E=E_1E_2\cdots E_N$ to $\sN(m)$. 
Then $\langle E,S_m \rangle = \langle E|_{\sN(m)},S_m|_{\sN(m)} \rangle$ for any $E$ and $S_m$.\footnote{
	For example, if $S_m=IXZ$ and $E=E_1E_2E_3$, then\\ \mbox{$E|_{\sN(m)}=(E_1E_2E_3)|_{\sN(m)} = E_2E_3$} and $S_m|_{\sN(m)}= XZ$. Then\\
	$\langle E,S_m \rangle = \langle E_1E_2E_3, IXZ \rangle = \left\langle E_2E_3, XZ \right\rangle = \langle E|_{\sN(m)},S_m|_{\sN(m)} \rangle.$
	}

A conventional BP$_4$ for decoding binary stabilizer codes is done as follows \cite{PC08}. 
In the initialization step, every variable node~$n$ passes the message
$\mbf{q}_{mn}=(q_{mn}^I,q_{mn}^X,q_{mn}^Y,q_{mn}^Z)$ $= {\mbf p}_n$ to every neighboring check node $m\in\sM(n)$. 
\begin{itemize} 
\item Horizontal Step:
At check node $m$, compute $\mbf{r}_{mn}=(r_{mn}^I,r_{mn}^X,r_{mn}^Y,r_{mn}^Z)$ and pass $\mbf{r}_{mn}$ as the message \mbox{$m\to n$} 
for every \mbox{$n\in\sN(m)$}, where $\forall~ W\in\{I,X,Y,Z\}$,
\beq{
r_{mn}^W = \sum_{E|_{\sN(m)}:~E_n=W, \atop \left\langle E|_{\sN(m)},S_m|_{\sN(m)} \right\rangle=z_m} \ps{ \prod_{n'\in\sN(m)\setminus n}q_{mn'}^{E_{n'}} }.
}
\item Vertical Step: 
At variable node $n$, compute $\mbf{q}_{mn}=(q_{mn}^I,q_{mn}^X,q_{mn}^Y,q_{mn}^Z)$ and pass $\mbf{q}_{mn}$ as the message \mbox{$n\to m$} 
for every $m\in\sM(n)$, where 
\beq{
q_{mn}^W = a_{mn}\, p_{n}^W \prod_{m'\in\sM(n)\setminus m} r_{m'n}^W 
}
with $a_{mn}$ a chosen scalar to let $\sum_{W\in\{I,X,Y,Z\}} q_{mn}^W = 1$.
\end{itemize} 
At variable node $n$, it also computes $q_n^W = p_{n}^W \prod_{m\in\sM(n)} r_{mn}^W $ 
to infer $\hat E_n = \argmax_{W\in\{I,X,Y,Z\}} q_n^{W}$.
The horizontal and vertical steps are iterated until an estimated $\hat{E}=\hat{E}_1\hat{E}_2\cdots \hat{E}_N$ is valid or a maximum number of iterations is reached.

\subsection{The refined BP$_4$ with scalar messages}

For a variable node, say variable node~1, and its neighboring check node $m$, we know from~(\ref{eq:syndrome_bit}) that
\begin{align*}
\langle E_1, S_{m1}\rangle &= z_m+\sum_{n=2}^N \langle E_n, S_{mn}\rangle \mod 2.  
\end{align*}
Given the value $\langle E_n, S_{mn}\rangle$ of (a possible) $E_n\in\{I,X,Y,Z\}$ and some $S_{mn}\in\{X,Y,Z\}$, we will know that $E_n$ commutes or anticommutes with $S_{mn}$, i.e.,  either $E_n\in \{I, S_{mn}\}$ or $E_n\in\{X,Y,Z\}\setminus S_{mn}$.
Consequently, the passed message should indicate more likely whether $E_n\in \{I, S_{mn}\}$ or $E_n\in\{X,Y,Z\}\setminus S_{mn}$.
In other words, the message from a neighboring check will tell us more likely whether the error $E_1$ commutes or anticommutes with $S_{m1}$. 
This suggests that a BP decoding of stabilizer codes with scalar messages is possible and we provide such an algorithm in Algorithm~\ref{alg:QBP}, referred to as \textit{parallel BP$_4$} (the same as \cite[Algorithm~3]{KL20}).
Note that the notation $r_{mn}^{(\langle W,S_{mn}\rangle)}$ is simplified as $r_{mn}^{\langle W,S_{mn}\rangle}$.

\begin{algorithm}
	\caption{:  $\dlt$-rule based quaternary BP decoding for binary stabilizer codes  	with a parallel schedule (parallel BP$_4$)} \label{alg:QBP}
	\textbf{Input}:  $S \in\{I,X,Y,Z\}^{M\times N}$, $z \in\{0,1\}^M$, and initial $\{(p_n^I, p_n^X, p_n^Y, p_n^Z)\}_{n=1}^N$.\\
	{\bf Initialization.}  For $n=1,2,\dots,N$ and $m\in\sM(n)$, let
		\begin{equation*}
		d_{mn}=q_{mn}^{(0)}-q_{mn}^{(1)}, 
		\end{equation*}
		\qquad where $q_{mn}^{(0)} = p_n^I+p_n^{S_{mn}}$ and $q_{mn}^{(1)} = 1 - q_{mn}^{(0)}$.
	
	{\bf Horizontal Step.} For $m=1,\dots,M$ and $n\in\sN(m)$, compute 
	\begin{equation*} \textstyle
	\dlt_{mn} = (-1)^{z_m}\prod_{n'\in\sN(m)\setminus n} \, d_{mn'}.   
	\end{equation*}
	
	{\bf Vertical Step.} For $n=1,2,\dots,N$ and $m\in\sM(n)$, do: 
	\begin{itemize}
		\item Compute $r_{mn}^{(0)} = (1+\dlt_{mn})/2, ~ r_{mn}^{(1)} = (1-\dlt_{mn})/2$,
		\begin{align*} 
		q_{mn}^I &= \textstyle p_n^I\prod_{m'\in\sM(n)\setminus m} \, r_{m'n}^{(0)} \text{ ~and } \\ 
		q_{mn}^W &= \textstyle p_n^W\prod_{m'\in\sM(n)\setminus m} \, r_{m'n}^{\langle W, S_{m'n}\rangle }, \mbox{ for } W\in\{X,Y,Z\}. 
		\end{align*}
		\item 
		Let ~$q_{mn}^{(0)} = a_{mn}(q_{mn}^I + q_{mn}^{S_{mn}})$ \\
		and ~$q_{mn}^{(1)} = a_{mn}(\sum_{W'\in\{X,Y,Z\}\setminus S_{mn}}q_{mn}^{W'})$,\\
		where $a_{mn}$ is a chosen scalar such that $q_{mn}^{(0)}+q_{mn}^{(1)}=1$.
		\item Update: $d_{mn} = q_{mn}^{(0)} - q_{mn}^{(1)}$. 
	\end{itemize}
	
	{\bf Hard Decision.} For $n=1,2,\dots,N$, compute
		\begin{align*}
		\qquad  q_n^I &= \textstyle p_n^I\prod_{m\in\sM(n)} \, r_{mn}^{(0)} \text{ ~and }  \\ 
				q_n^W &= \textstyle p_n^W\prod_{m\in\sM(n)} \, r_{mn}^{\langle W,S_{mn}\rangle}, \mbox{ for } W\in\{X,Y,Z\}. 
		\end{align*} 
		\begin{itemize}
		\item[]	Let $\hat E_n = \argmax_{W\in\{I,X,Y,Z\}} q_n^{W}$.		
		\end{itemize}

	\begin{itemize}
		\item Let $\hat E = \hat E_1\hat E_2\cdots\hat E_N$. 
		\begin{itemize}
			\item If $\langle \hat E, S_m \rangle = z_m ~\forall~ m$, halt   and return ``SUCCESS'';
			\item otherwise, if a maximum number of iterations is reached, halt   and return ``FAIL'';
			\item otherwise, repeat from the horizontal step.
		\end{itemize}
	\end{itemize}
\end{algorithm}

It can be shown that Algorithm~\ref{alg:QBP} has exactly the same output as the conventional BP$_4$ (as outlined in Sec.~\ref{sec:Simp_BP4}). 
In particular, Algorithm~\ref{alg:QBP} has a check-node efficiency 16-fold improved from the conventional BP$_4$. 

Similar to the classical case, we can consider running Algorithm~\ref{alg:QBP} with a serial schedule,
referred to as \emph{serial BP$_4$} (cf. \cite[Algorithm~4]{KL20}).
Using serial BP$_4$ may prevent the decoding oscillation (as an example based on the $[[5,1,3]]$ code in \cite[Fig.\,7]{KL20}) 
and improve the empirical performance (e.g., for decoding a hypergraph-product code \cite[Fig.\,8]{KL20}).

\section{Simulation Results} \label{sec:NorQBP}


\begin{algorithm}[htbp] \caption{: Normalized BP$_4$ with a given parameter $\alpha_v$} \label{alg:QBP-av}
Identical to Algorithm~\ref{alg:QBP}, except that  
$q_{mn}^{(0)}$ and $q_{mn}^{(1)}$ are respectively replaced by 
$q_{mn}^{(0)}=a_{mn}(q_{mn}^I+q_{mn}^{S_{mn}})^{1/\alpha_v}$ and $q_{mn}^{(1)}=a_{mn}(\sum_{W'\in\{X,Y,Z\}\setminus S_{mn}}q_{mn}^{W'})^{1/\alpha_v}$ for some $\alpha_v>0$, where $a_{mn}$ is a chosen scalar such that $q_{mn}^{(0)}+q_{mn}^{(1)}=1$.
\end{algorithm}

\begin{figure}[htbp]
	\centering \includegraphics[width=0.5\textwidth]{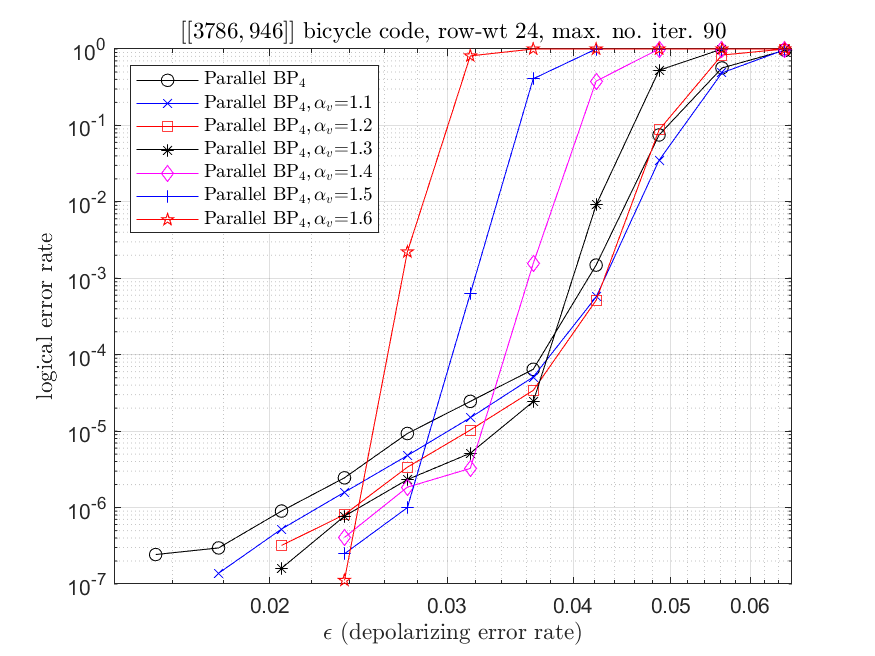}
	\caption{Performance of decoding the $[[3786,946]]$ bicycle code.}			\label{fig:3786_Hk24_av}	\vspace*{\floatsep}
	\centering \includegraphics[width=0.5\textwidth]{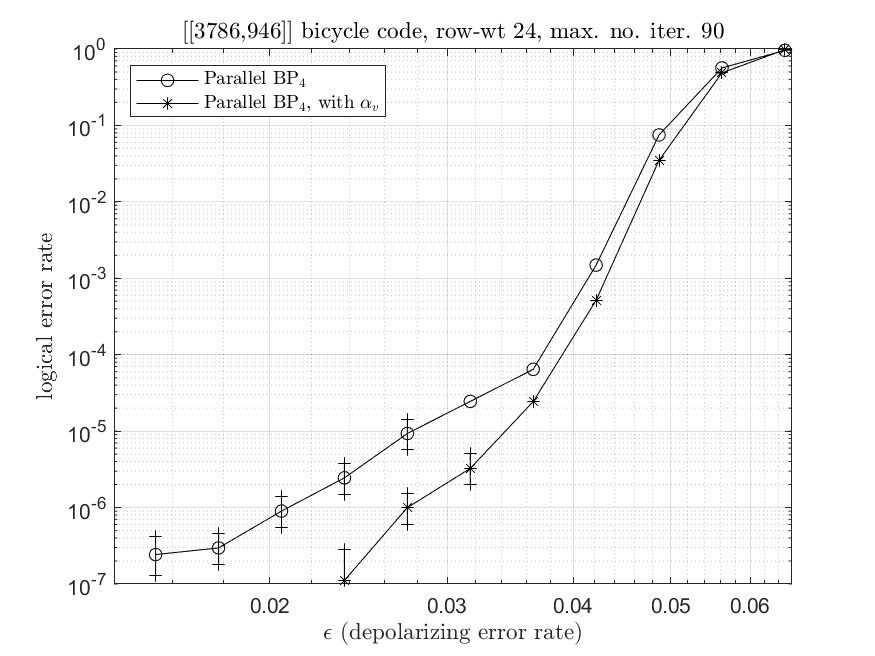}
	\caption{Performance of decoding the $[[3786,946]]$ bicycle code with different $\alpha_v$ used for different $\epsilon$.
		This figure is derived from the data in Fig.\,\ref{fig:3786_Hk24_av}.}	\label{fig:3786_Hk24_opt_av}
	\end{figure}

Having scalar messages 
allows us to apply the {\it message normalization} \cite{CF02b,CDE+05,YHB04}.
This suppresses the wrong belief looped in the short cycles (due to sub-matrices like $\bsmtx{X&Y\\Z&Z}$ in Fig.\,\ref{fig:S2x3}).
The message normalization can be applied to $(r_{mn}^{(0)},r_{mn}^{(1)})$ \cite[Algorithm~5]{KL20} or $(q_{mn}^{(0)},q_{mn}^{(1)})$ \cite[Algorithm~6]{KL20}.
Many simulation results were provided in \cite{KL20}, and normalizing $(q_{mn}^{(0)},q_{mn}^{(1)})$ has a lower error-floor.
We restate \cite[Algorithm~6]{KL20} as Algorithm~\ref{alg:QBP-av} here (where $(\cdot)^{1/\alpha_v}$ can be efficiently done by one multiplication and two additions \cite{Sch99}). Some more results will be provided. 
We try to collect at least 100 logical errors for every data-point; or otherwise an error bar between two crosses shows a 95\% confidence interval (omitted in Fig.\,\ref{fig:3786_Hk24_av}).

In the simulation, we do not restrict the measurements to be local.
We consider bicycle codes (a kind of random sparse codes) \cite{MMM04} since they have small ancilla overhead and good performance, benchmarked by code rate vs.~physical error rate \cite[Fig.\,10]{MMM04}.\footnote{
	Using random sparse codes needs long qubit-connectivity, which is possibly supported by trapped-ions or nuclear-spins \cite{CTV17}.
	If the measurements must be local, then topological codes are usually considered (with another benchmark called threshold usually used).
	Our proposed algorithm can also decode topological codes but needs some extension to have a good performance. This will be addressed in another our manuscript under preparation.
	}
But BP decoding of this kind of codes have some error-floor subject to improve \cite[Fig.\,6]{MMM04}.

\begin{figure}[htbp]
	\centering \includegraphics[width=0.5\textwidth]{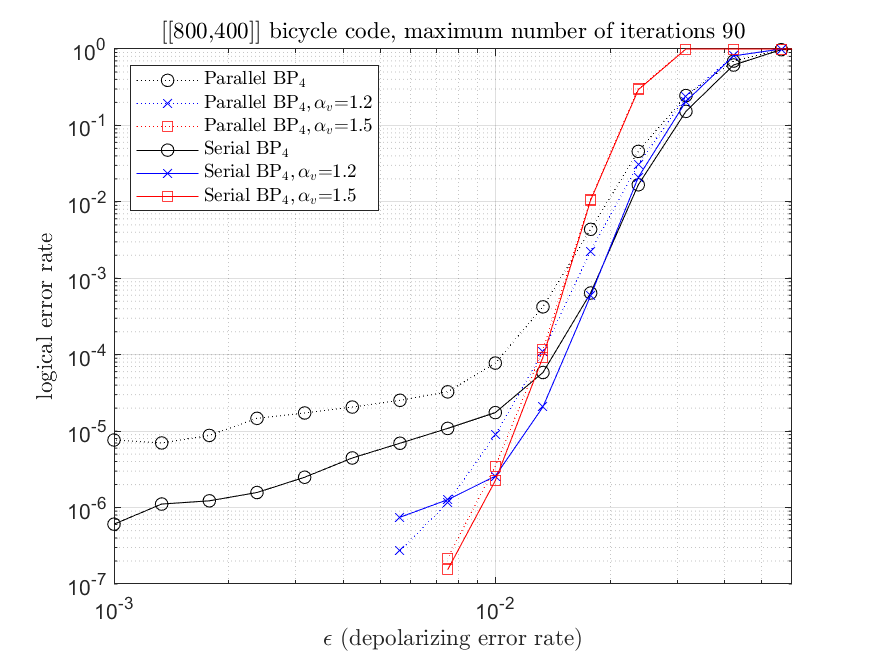}
	\caption{Performance of decoding a random $[[800,400]]$ bicycle code in \cite{KL20}.} 						\label{fig:800_av}			\vspace*{\floatsep}
	\centering \includegraphics[width=0.5\textwidth]{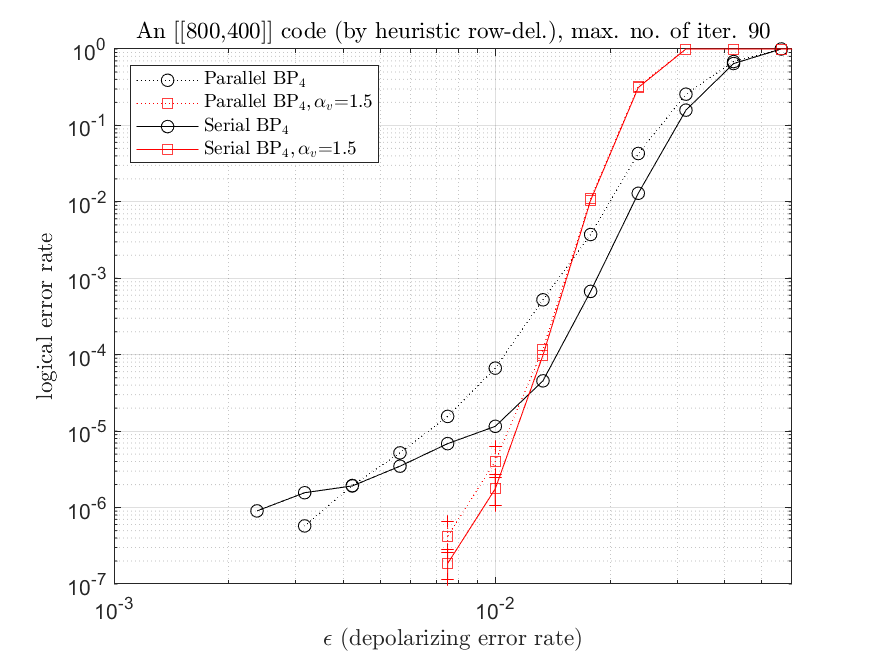}
	\caption{Performance of decoding another $[[800,400]]$ bicycle code constructed by the heuristic approach (resulting in more uniform column-weights).}	\label{fig:800_Hjmin_av}	
	\end{figure}

First, a random bicycle code is easy to construct \cite{PC08} but could have a high error-floor \cite{KL20}.
For convenience, we replot \cite[Fig.\,16]{KL20} as Fig.\,\ref{fig:800_av} here. 
Observe the high error-floor before using message normalization. Using $\alpha_v$ improves the error-floor a lot.
We implemented an overflow (underflow) protection in BP as in \cite{Mac99} and confirmed that the original error-floor is not due to numerical overflow/underflow.
It should be due to the random construction (especially that a row-deletion step is performed randomly). 
MacKay, Mitchison, and McFadden suggested to use a heuristic approach to do the construction (to have more uniform column-weights in the check matrix after the row-deletion step) \cite{MMM04}.\footnote{
	Except for special irregular designs, BP converges more well when the column-weights are more uniform \cite{Gal63,Mac99,MMM04}.
	We consider to either minimize the column-weight variance ({\sc min-var} approach) 
	or minimize the difference of the largest and smallest column-weights ({\sc min-max} approach).
	The \mbox{\sc min-var} approach is usually better;
	but if many rows are deleted (for a larger code rate), the {\sc min-max} approach could be better.
	The $[[800,400]]$ code (rate~1/2) constructed here is by the {\sc min-max} approach. 
	The $[[3786,946]]$ code (rate~1/4) constructed here is by the {\sc min-var} approach. 
	}

An $[[800,400]]$ bicycle code is constructed by the heuristic approach, with the other parameters the same as the code in Fig.\,\ref{fig:800_av}.
The performance of this new code is shown in Fig.\,\ref{fig:800_Hjmin_av}, where
the BP$_4$ error-floor (without message normalization) is improved compare to Fig.\,\ref{fig:800_av}.
However, using $\alpha_v=1.5$ further lowers the error-floor to a level like the improved case in Fig.\,\ref{fig:800_av}. 
The two figures show that running (scalar-based) BP$_4$ with message normalization has a more robust performance. 
This provides more flexibility in code construction (for if some rows (measurements) must be kept or deleted). 

Next we construct a $[[3786,946]]$ bicycle code with row-weight 24, using the heuristic approach. 
A code with such parameters has some error-floor for a block error rate $<10^{-4}$, when decoded by BP$_2$ in the bit-flip channel \cite[Fig.\,6]{MMM04}.
We decode the constructed code by BP$_4$ in the depolarizing channel, and plot the decoding performance in Fig.\,\ref{fig:3786_Hk24_av}.
%
There is a similar error-floor for a logical error rate $<10^{-4}$ before using message normalization.
After using $\alpha_v$, the error-floor performance is much improved.
A smaller $\epsilon$ usually needs a larger $\alpha_v$ for suppression (since $\mbf p_n$ is quite biased with a large $p_n^I=1-\epsilon$).
As a reference, we derive Fig.\,\ref{fig:3786_Hk24_opt_av} from Fig.\,\ref{fig:3786_Hk24_av} by using different $\alpha_v$ for different $\epsilon$ to see the improvement.

\section{Conclusion} \label{sec:Conclu}

We proposed a $\dlt$-rule based BP$_4$ for decoding quantum stabilizer codes by passing scalar messages.
This is achieved by exploiting the binary property of the syndrome.
The proposed BP$_4$ is equivalent to the conventional BP$_4$ but with a check-node efficiency 16-fold improved.
%
The message normalization can be naturally applied to the scalar messages.
We performed the simulation by decoding several bicycle codes. 
Different decoding schedules were also considered.
Using the proposed BP$_4$ with message normalization and a suitable schedule results in a significantly improved performance and error-floor. 



%

\bibliographystyle{IEEEtran}
\bibliography{References}

\end{document}